\documentclass[11pt, oneside]{article} 
\usepackage[utf8]{inputenc}
\usepackage[utf8]{inputenc}
\usepackage{geometry}                		
\usepackage{graphicx}				
\usepackage{amssymb}
\usepackage{amsmath}
\usepackage{subfig}
\usepackage{caption}
\usepackage{comment}
\usepackage{amsthm,verbatim,amsfonts,amscd,float,physics}
\usepackage{appendix}

\usepackage[utf8]{inputenc}
\usepackage{xcolor}
\numberwithin{equation}{section}
\usepackage{titling, lipsum}
\title{A Matrix Big Bang on a Quantum Computer }
\author{ Viti Chandra$^{(1)}$, Yuan Feng$^{(2)}$, Michael McGuigan$^{(3)}$\\
(1) Half Hollow Hills High School West \\
(2) University of California, Berkeley\\
(3) email contact: michael.d.mcguigan@gmail.com
}
\date{}

\begin{document}
\begin{titlingpage}

\maketitle
\begin{abstract}
M-theory is a mysterious theory that seeks to unite different string theories in one lower dimension. The most studied example is eleven dimensional but other dimensions have been considered. The non-critical M-theories seek to unite different non-critical string theories. From the point of view of computing, non-critical M-theories should be simpler to simulate as they have fewer fields than eleven dimensional M-theory. For example eleven dimensional M-theory can be simulated using nine bosonic fields and a Hamiltonian derived from the dimensional reduction of 10d Super-Yang Mills to 0+1 dimensions. Non-critical M-theory in four dimensions on the other hand has two bosonic fields derived from the dimensional reduction of 3d Super-Yang-Mills to 0+1 dimensions. Non-critical M-theory in three dimensions is even simpler  with a single bosonic field derived from the dimensional reduction of 2d Super-Yang-Mills to 0+1 dimensions. The simplicity of non-critical M-theory carries over to quantum computing and we show that the quantum simulation requires fewer qubits and Pauli terms than critical M-theory. As an example quantum calculation we study the quantum computation of the ground state energy of Matrix models of non-critical M-theory in 3d in the finite difference and  oscillator basis and compare the accuracy, number of qubits and number of Pauli terms of the different basis using the Variational Quantum Eigensolver (VQE) algorithm. We study non-critical M- Theory solutions with space-time singularities referred to as a "Matrix Big Bang" on the Quantum Computer using the Evolution of Hamiltonian (EOH) quantum algorithm using the Trotter approximation and compare the accuracy and results the can be obtained using quantum computation. Finally we consider the BRST quantization of the 3d M-theory Matrix model using quantum computation and compute BRST invariant states by studying the BRST Laplacian using the VQE algorithm.

\end{abstract}
\end{titlingpage}

\section{Introduction}

Quantum computing is a potentially disruptive computing paradigm with scientific applications such as quantum chemistry, nuclear physics and high energy physics\cite{Alexeev:2020xrq}
\cite{Mihalikova:2021tts}
\cite{Tilly:2021jem}
\cite{Cloet}
\cite{Bauer:2022hpo}\cite{Faulkner:2022mlp}
\cite{Gharibyan:2020bab}. In \cite{Rinaldi:2021jbg} quantum computing was applied to the study of Matrix models. Recently quantum computing applied to SYK models have been used to simulate wormholes in a dual gravity model \cite{Lykken}.  Matrix models can be used to formulate non-perturbative approaches to String Theory \cite{Dijkgraaf:1997vv}
\cite{Motl:1997th} and can be used to formulate M-theory \cite{Witten:1995ex}\cite{Banks:1996vh} in certain backgrounds such as the pp-wave geometry\cite{Berenstein:2002jq}. Another quantum computing application of Matrix models is to nuclear physics as they are much simpler to simulate on near term quantum computers that gauge field theories \cite{Miceli:2019snu}
\cite{Feng:2021mdi}
\cite{Butt:2022xyn}. Although perturbative String theory is a finite approach to graviton scattering there are still open issues with singularities in cosmology and black holes which are unaddressed in perturbative string theory \cite{Horowitz:1989bv} and Matrix theory may be an approach to these problems. For example in \cite{Craps:2005wd}
\cite{Craps:2010cn}
\cite{Chen:2005mga}
\cite{Chen:2006rm} a Matrix Big Bang solution was studied. In this paper we will study the implementation of a simpler version of 11d M-theory called 3d M-theory which can be more easily realized on today's quantum computers and simulators.  

This paper is organized as follows. In section two we discuss 3d M-theory and emphasize the Matrix version of the theory. I section three we apply the Variational Quantum Eigensolver algorithm applied to 3d M-theory. In section four we apply the Evolution of Hamiltonian quantum algorithm to a time dependent Matrix Model associated a subspace of the Matrix Big Bang space-time. In section four we discuss a BRST approach to the quantum computing of Matrix models of 3d M-theory. In section five we list the main conclusions and outlook for extensions of the simulations to more general backgrounds.   

\section{3d M-theory}

3d M-theory is a simplified version of M-theory which usually exists in 11d. The usual form of M-theory can be defined through a Matrix theory which contains nine bosonic fields and is obtained by the dimensional reduction of 10d Super yang-Mills to 0+1 dimensions. 3d M-theory can be defined similarly through a Matrix theory with a single bosonic field obtained from the dimensional reduction of 2d Super Yang-Mills to 0+1 dimensions. In addition there are other ways to define the 3d M-theory in terms of Supermembranes and fermionic field theories \cite{McGuigan:2004sq}
\cite{Horava:2005tt}
\cite{Horava:2005wm}
\cite{Horava:2007ds}\cite{Petkou:2005se}. In this paper we consider the matrix formulation considered in \cite{Park:2005pz}
\cite{Kim:2006wg}. The Lagrangian for the $SU(N)$ Matrix model for 3d M-theory theory is given by:
\[L = Tr\left[ {\frac{1}{2}{{\left( {{D_t}X} \right)}^2} + i\frac{1}{2}\psi^\dag {D_t}\psi  + g X\psi \psi  - \frac{1}{2}{X^2}- \psi^\dag \psi } \right]\]
from which one can obtain the Hamiltonian. 
\[H = \frac{1}{2}{p_a}{p_a} + \frac{1}{2}{x_a}{x_a} + \psi_a^\dag \psi_a+ g{\varepsilon _{abc}}\psi _a {x_b}{\psi _c}\]
where we have expanded $P$ and $X$ is terms of Pauli matrices with coefficients $p_a$ and $x_a$ for the simplest case of $SU(2)$. Here $g$ is the coupling constant of the theory with the quadratic term representing a mass deformation.

\subsection*{Quantum computation}

To represent the Matrix theory for 3d M-theory on a quantum computer one must perform the Hamiltonian mapping which represents the Hamiltonian as a finite matrix using a basis and expands the matrix in terms of Pauli terms. Once one has the Hamiltonian represented in terms of qubits and Pauli terms one can run various quantum algorithms with their different quantum circuits on a quantum computer or quantum computer simulator. The two algorithms we will consider in this paper are the Variational Quantum Eigensolver (VQE) and Evolution of Hamiltonian (EOH) quantum algorithms. 

\subsection*{Gaussian or Simple Harmonic Oscillator basis}

For quantum computation  one needs a basis for the Hamiltonian. 
The Gaussian or Harmonic Oscillator representation or basis is derived using the matrix treatment of the simple harmonic oscillator. The position operator is given by:
\begin{equation} 
 Q_{osc} = \frac{1}{\sqrt{2}}\begin{bmatrix}
 
   0 & {\sqrt 1 } & 0 &  \cdots  & 0  \\ 
   {\sqrt 1 } & 0 & {\sqrt 2 } &  \cdots  & 0  \\ 
   0 & {\sqrt 2 } &  \ddots  &  \ddots  & 0  \\ 
   0 & 0 &  \ddots  & 0 & {\sqrt {N-1} }  \\ 
   0 & 0 &  \cdots  & {\sqrt {N-1} } & 0  \\ 
\end{bmatrix}
  \end{equation}
and the momentum operator is:
\begin{equation}
 P_{osc} = \frac{i}{\sqrt{2}}\begin{bmatrix}
 
   0 & -{\sqrt 1 } & 0 &  \cdots  & 0  \\ 
   {\sqrt 1 } & 0 & -{\sqrt 2 } &  \cdots  & 0  \\ 
   0 & {\sqrt 2 } &  \ddots  &  \ddots  & 0  \\ 
   0 & 0 &  \ddots  & 0 & -{\sqrt {N-1} }  \\ 
   0 & 0 &  \cdots  & {\sqrt {N-1} } & 0  \\ 
\end{bmatrix}
  \end{equation}
The Hamiltonian is then $H(q,p) = H( Q_{osc}, P_{osc})$ and is expanded in terms of qubits and Pauli terms.

\subsection*{Finite difference basis}

 For the finite difference basis the position matrix is diagonal:
\begin{equation}{\left( {{Q_{fd}}} \right)_{j,k}} = \sqrt {\frac{1}{{2N}}} (2j - (N + 1)){\delta _{j,k}}\end{equation}
and the momentum-squared matrix is:
\begin{equation} 
 P_{fd}^2 = \frac{N}{2}\begin{bmatrix}
 
   2 & - 1  & 0 &  \cdots  & 0  \\ 
   -1 & 2 & -1 &  \cdots  & 0  \\ 
   0 & -1 &  \ddots  &  \ddots  & 0  \\ 
   0 & 0 &  \ddots  & 2 & -1  \\ 
   0 & 0 &  \cdots  & -1 & 2  \\ 
\end{bmatrix}
  \end{equation}
The Hamiltonian is then represented as $H(q,p) = H( Q_{fd}, P_{fd})$

\subsection*{Construction of Hamiltonians}

\subsubsection*{Oscillator basis}
To construct the 3d M-theory Hamiltonian is the oscillator basis we define the three bosonic Matrices:
\[{a_1} = \left( \begin{array}{l}
\begin{array}{*{20}{c}}
0&1\\
0&0
\end{array}\begin{array}{*{20}{c}}
0&0\\
{\sqrt 2 }&0
\end{array}\\
\begin{array}{*{20}{c}}
0&0\\
0&0
\end{array}\begin{array}{*{20}{c}}
0&{\sqrt 3 }\\
0&0
\end{array}
\end{array} \right) \otimes \left( \begin{array}{l}
\begin{array}{*{20}{c}}
1&0\\
0&1
\end{array}\begin{array}{*{20}{c}}
0&0\\
0&0
\end{array}\\
\begin{array}{*{20}{c}}
0&0\\
0&0
\end{array}\begin{array}{*{20}{c}}
1&0\\
0&1
\end{array}
\end{array} \right) \otimes \left( \begin{array}{l}
\begin{array}{*{20}{c}}
1&0\\
0&1
\end{array}\begin{array}{*{20}{c}}
0&0\\
0&0
\end{array}\\
\begin{array}{*{20}{c}}
0&0\\
0&0
\end{array}\begin{array}{*{20}{c}}
1&0\\
0&1
\end{array}
\end{array} \right)\]
\[{a_2} = \left( \begin{array}{l}
\begin{array}{*{20}{c}}
1&0\\
0&1
\end{array}\begin{array}{*{20}{c}}
0&0\\
0&0
\end{array}\\
\begin{array}{*{20}{c}}
0&0\\
0&0
\end{array}\begin{array}{*{20}{c}}
1&0\\
0&1
\end{array}
\end{array} \right) \otimes \left( \begin{array}{l}
\begin{array}{*{20}{c}}
0&1\\
0&0
\end{array}\begin{array}{*{20}{c}}
0&0\\
{\sqrt 2 }&0
\end{array}\\
\begin{array}{*{20}{c}}
0&0\\
0&0
\end{array}\begin{array}{*{20}{c}}
0&{\sqrt 3 }\\
0&0
\end{array}
\end{array} \right) \otimes \left( \begin{array}{l}
\begin{array}{*{20}{c}}
1&0\\
0&1
\end{array}\begin{array}{*{20}{c}}
0&0\\
0&0
\end{array}\\
\begin{array}{*{20}{c}}
0&0\\
0&0
\end{array}\begin{array}{*{20}{c}}
1&0\\
0&1
\end{array}
\end{array} \right)\]
\[{a_3} = \left( \begin{array}{l}
\begin{array}{*{20}{c}}
1&0\\
0&1
\end{array}\begin{array}{*{20}{c}}
0&0\\
0&0
\end{array}\\
\begin{array}{*{20}{c}}
0&0\\
0&0
\end{array}\begin{array}{*{20}{c}}
1&0\\
0&1
\end{array}
\end{array} \right) \otimes \left( \begin{array}{l}
\begin{array}{*{20}{c}}
1&0\\
0&1
\end{array}\begin{array}{*{20}{c}}
0&0\\
0&0
\end{array}\\
\begin{array}{*{20}{c}}
0&0\\
0&0
\end{array}\begin{array}{*{20}{c}}
1&0\\
0&1
\end{array}
\end{array} \right) \otimes \left( \begin{array}{l}
\begin{array}{*{20}{c}}
0&1\\
0&0
\end{array}\begin{array}{*{20}{c}}
0&0\\
{\sqrt 2 }&0
\end{array}\\
\begin{array}{*{20}{c}}
0&0\\
0&0
\end{array}\begin{array}{*{20}{c}}
0&{\sqrt 3 }\\
0&0
\end{array}
\end{array} \right)\]
\begin{equation}
\end{equation}
and three fermion matrices
\[{c_1} = \left( {\begin{array}{*{20}{c}}
0&1\\
0&0
\end{array}} \right) \otimes \left( {\begin{array}{*{20}{c}}
1&0\\
0&1
\end{array}} \right) \otimes \left( {\begin{array}{*{20}{c}}
1&0\\
0&1
\end{array}} \right)\]
\[{c_2} = \left( {\begin{array}{*{20}{c}}
1&0\\
0&1
\end{array}} \right) \otimes \left( {\begin{array}{*{20}{c}}
0&1\\
0&0
\end{array}} \right) \otimes \left( {\begin{array}{*{20}{c}}
1&0\\
0&1
\end{array}} \right)\]
\[{c_3} = \left( {\begin{array}{*{20}{c}}
1&0\\
0&1
\end{array}} \right) \otimes \left( {\begin{array}{*{20}{c}}
1&0\\
0&1
\end{array}} \right) \otimes \left( {\begin{array}{*{20}{c}}
0&1\\
0&0
\end{array}} \right)\]
\begin{equation}
\end{equation}
One then defines 
\[{A_i} = {a_i} \otimes {I_8}\]
\begin{equation}{C_i} = {I_{64}} \otimes c\end{equation}
where $I_d$ is the $d\times d$ identity matrix. The matrix Hamiltonian is then given by:
\begin{equation}H = {H_2} + {H_{123}} + {H_{231}} + {H_{312}}\end{equation}
where
\[{H_2} = A_1^\dag {A_1} + A_2^\dag {A_2} + A_3^\dag {A_3} + C_1^\dag {C_1} + C_2^\dag {C_2} + C_3^\dag {C_3}\]
\[{H_{123}} = \frac{g}{{\sqrt 2 }}\left( {{C_1}{C_2} + {{\left( {{C_1}{C_2}} \right)}^\dag }} \right)\left( {{A_3} + A_3^\dag } \right)\]
\[{H_{231}} = \frac{g}{{\sqrt 2 }}\left( {{C_2}{C_3} + {{\left( {{C_2}{C_3}} \right)}^\dag }} \right)\left( {{A_1} + A_1^\dag } \right)\]
\begin{equation}{H_{312}} = \frac{g}{{\sqrt 2 }}\left( {{C_3}{C_1} + {{\left( {{C_3}{C_1}} \right)}^\dag }} \right)\left( {{A_2} + A_2^\dag } \right)\end{equation}

\subsubsection*{Finite difference basis}

To construct the Matrix theory Hamiltonian in the finite difference basis we define two Matrices through (). In our calculations we set $N=4$
\[P_{fd}^2 = \frac{4}{2}\left( {\begin{array}{*{20}{c}}
2&{ - 1}&0&0\\
{ - 1}&2&{ - 1}&0\\
0&{ - 1}&2&{ - 1}\\
0&0&{ - 1}&2
\end{array}} \right)\]
\[{X_{fd}} = \sqrt {\frac{1}{{2(4)}}} \left( {\begin{array}{*{20}{c}}
{ - 3}&0&0&0\\
0&{ - 1}&0&0\\
0&0&1&0\\
0&0&0&3
\end{array}} \right)\]
Now define:
\[P_1^2 = P_{fd}^2 \otimes {I_4} \otimes {I_4} \otimes {I_8}\]
\[P_2^2 = {I_4} \otimes P_{fd}^2 \otimes {I_4} \otimes {I_8}\]
\[P_3^2 = {I_4} \otimes {I_4} \otimes P_{fd}^2 \otimes {I_8}\]
\[{X_1} = {X_{fd}} \otimes {I_4} \otimes {I_4} \otimes {I_8}\]
\[{X_2} = {I_4} \otimes {X_{fd}} \otimes {I_4} \otimes {I_8}\]
\begin{equation}{X_3} = {I_4} \otimes {I_4} \otimes {X_{fd}} \otimes {I_8}\end{equation}
The fermions are defined the same as in the oscillator basis. The Matrix Hamiltonian in the finite difference basis is then:
\begin{equation}{H_{fd}} = H_{fd}^{(2)} + H_{fd}^{(3)}\end{equation}
where
\[H_{fd}^{(2)} = \frac{1}{2}\left( {P_1^2 + P_2^2 + P_3^2 +X_1^2 + X_2^2 + X_3^2  } \right) + C_1^\dag {C_1} + C_2^\dag {C_2} + C_3^\dag {C_3} - \frac{3}{2}{I_{512}}\]
\begin{equation}H_{fd}^{(3)} = g\left( {\left( {{C_1}{C_2} + {{\left( {{C_1}{C_2}} \right)}^\dag }} \right){X_3} + \left( {{C_2}{C_3} + {{\left( {{C_2}{C_3}} \right)}^\dag }} \right){X_1} + \left( {{C_3}{C_1} + {{\left( {{C_3}{C_1}} \right)}^\dag }} \right){X_2}} \right)\end{equation}
To represent either Hamiltonian in a quantum algorithm it is necessary to represent it in terms of an expansion of Pauli terms which are tensor product combinations of $I_2$ and the three Pauli matrices. Although there are potentially $4^{N_q}$ such Pauli terms for $N_q$ qubits we find that there are substantially fewer terms in the expansion that occur in practice in either basis.

\subsubsection*{VQE results}

The hybrid quantum algorithm we used was the Variational Quantum Eigensolver (VQE). This algorithm represents a trial variational wave function using a quantum circuit containing rotation gates parametrized by angular parameters $\theta_i$. The quantum circuit is evaluated on the quantum computer or quantum computer simulator and the angles are varied to minimize:
\begin{equation}E(\theta ) = \frac{{\left\langle {\psi ({\theta _i})} \right|H\left| {\psi ({\theta _i}} \right\rangle }}{{\left\langle {\psi ({\theta _i})} \right|\left. {\psi ({\theta _i})} \right\rangle }}\end{equation}
with respect to $\theta_i$ using an optimiser which runs on a classical computer. For our calculations we used the Sequential Least SQuares Programming (SLSQP) optimzer, and the variational ansatz described by a quantum circuit with $R_y$ rotation blocks and $C_z$ entanglement blocks. We used nine qubits so our Hamiltonians using both oscillator and finite difference basis were represented as $515 \times 512$ matrices. Our results are recorded in table 1 for a coupling constant value of $g=.1$.
\begin{figure}
\centering
  \includegraphics[width = 1 \linewidth]{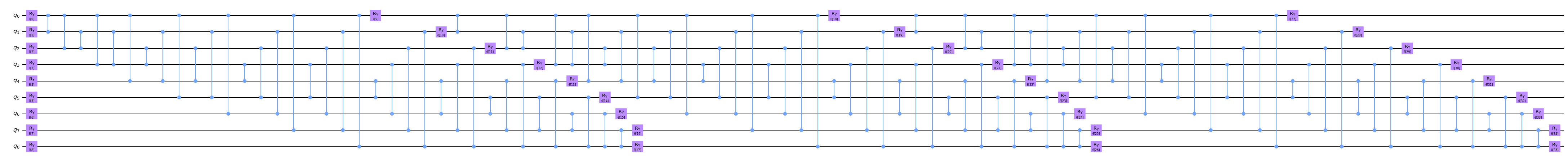}
  \caption{Variational ansatz represented as a 9 qubit quantum circuit for the quantum VQE computation of the ground state of the Hamiltonian for the 3d M-theory Matrix Model using the oscillator basis. The ansatz uses  parametrized $R_y$  gates with a depth of three and thirty six parameters }
  \label{fig:Radion Potential}
\end{figure}
\begin{figure}[!htb]
\centering
\minipage{0.5\textwidth}
  \includegraphics[width=\linewidth]{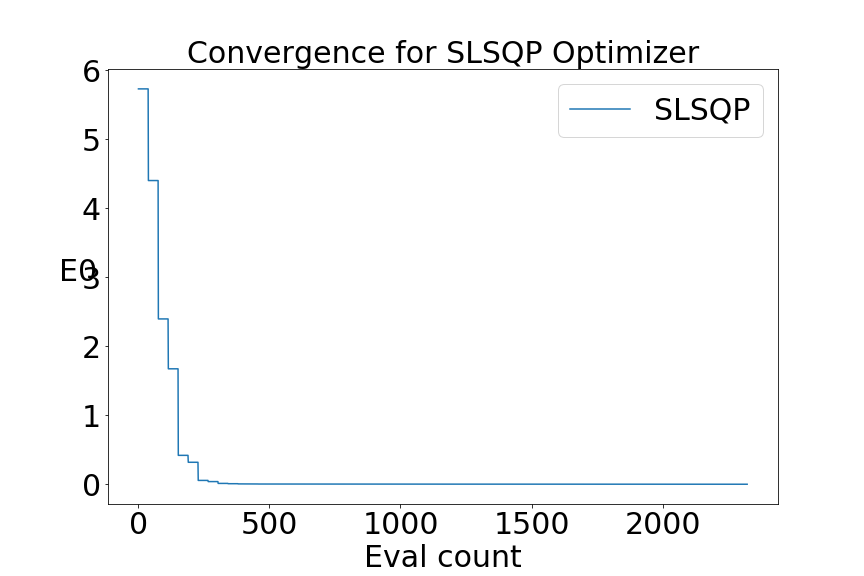}
\endminipage\hfill
\minipage{0.5\textwidth}
  \includegraphics[width=\linewidth]{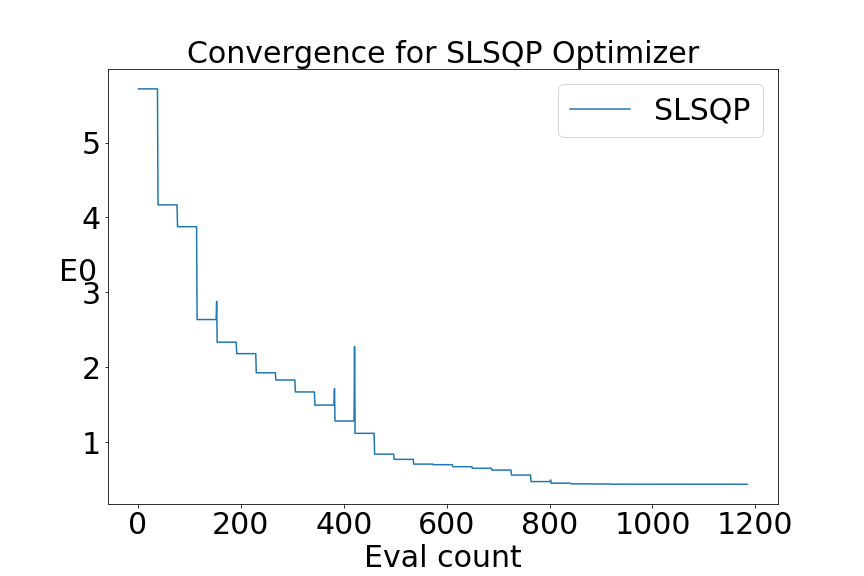}
\endminipage\hfill
\caption{Convergence graph for the quantum VQE computation of the lowest Energy eigenstate for the 3d M-theory Matrax Model with $g=.1$ using the oscillator basis (left) and finite difference basis (right). 
}
\end{figure}

\begin{table}[ht]
\centering
\begin{tabular}{|l|l|l|l|l|l|l|}
\hline
3d M-theory $g = .1$  & basis    & Qubits  &  Paulis  & Exact Result & Exact Discrete & VQE Result \\ \hline
$E_0$ with $g=.1$  & osc & 9 & 35  & $0.0$ &  $-0.005$ & $-0.00167$ \\ \hline
$E_0$ with $g=.1$ & fd & 9 & 28  & $0.0$ &  $ 0.18447904$ & $ 0.43679594$ \\ \hline

\end{tabular}
\caption{\label{tab:BasisCompare}  VQE results for the lowest eigenvalue of the Hamiltonian for 3D M-theory using the oscillator and finite difference  basis. 
}
\end{table}
The results indicate that the oscillator basis is more accurate than the finite difference basis for VQE computations of Matrix theory at least for 9 qubits. More qubits may be necessary  to accurately represent the Hamiltonian using the finite difference basis. Nevertheless we shall find the finite difference basis very useful for the EOH algorithm which we study in the next section. This is because the kinetic term in the Matrix Hamiltonian can be mapped to a graph Laplacian \cite{Feng:2021whf} and can be used to represent the computations of the transition probability of Matrix theory in terms of the quantum walk quantum algorithm on the graph.


\section{Matrix cosmology}

The 3d Matrix Lagrangian for general time dependent deformations $\Lambda(t)$ and $\rho(t)$ is given by \cite{Park:2005pz}
\cite{Kim:2006wg}:
\begin{equation}L = Tr\left[ {\frac{1}{2}{{\left( {{D_t}X} \right)}^2} + i\frac{1}{2}\psi {D_t}\psi  + X\psi \psi  + \frac{1}{2}\Lambda (t){X^2} + \rho (t)X} \right]\end{equation}
For the space-time metric associated with the usual 11d M-theory given by \cite{Craps:2010cn}:
\begin{equation}d{s^2} =  - 2d\tau dv - \frac{{{\alpha ^2}{{\left( {{Z^i}} \right)}^2} - \left( {{\beta ^2} - 2\alpha \beta } \right){{\left( {{Z^{11}}} \right)}^2}}}{{{{\left( {2\alpha \tau } \right)}^2}}}d{\tau ^2} + {\left( {d{Z^i}} \right)^2} + {\left( {d{Z^{11}}} \right)^2}\end{equation}
the bosonic and fermionic Matrix Lagragians are:
\begin{align}
&{L_B} =  Tr\left\{ {\frac{1}{{2R}}\left[ {{{\left( {{D_\tau }{Z^i}} \right)}^2} + {{\left( {{D_\tau }{Z^{11}}} \right)}^2}} \right] - \frac{R}{2}\left[ {{{\left[ {{Z^i},{Z^j}} \right]}^2} + 2{{\left[ {{Z^i},{Z^{11}}} \right]}^2}} \right] - \frac{{{\alpha ^2}{{\left( {{Z^i}} \right)}^2} - \left( {{\beta ^2} - 2\alpha \beta } \right){{\left( {{Z^{11}}} \right)}^2}}}{{{{\left( {2\alpha \tau } \right)}^2}}}} \right\} \\ \nonumber
&{L_F} = i{\theta ^T}{D_\tau }\theta  - R{\theta ^T}{\gamma _i}\left[ {{Z^i},\theta } \right] - R{\theta ^T}{\gamma _{11}}\left[ {{Z^{11}},\theta } \right]
\end{align}
To make contact with the 3d Matrix theory we fix the hypersurface $Z^i = 0$ and set $Z^{11}= X$. Then the deformation $\Lambda(t)$ is
\begin{equation}\Lambda (t) = \frac{{\left( {{\beta ^2} - 2\alpha \beta } \right)}}{{{{\left( {2\alpha t} \right)}^2}}}\end{equation}
where we have set $t=\tau$.

To study this Matrix theory on a quantum computer we use the Evolution of Hamiltonian (EOH) algorithm. This algorithm computes the transition probability for time evolution for a Hamiltonian. In our case the Hamiltonian is time dependent so we have to break up the evolution in several time steps and compose the Unitary evolution between time steps and compose each of theses to get the final state of the evolution. We study the Matrix Hamiltonian for the simplest case of $SU(2)$ and use the finite difference basis. The evolution is the product of three terms corresponding to each component of X expanded in terms the SU(2) Lie algebra so we can concentrate on one of them to get the other two.

The time evolution for the bosonic Hamiltonian is exactly soluble and is generalization of the time independent simple harmonic oscillator time evolution amplitude \cite{Mukhanov:2007zz}. If one has two independent solutions $u(t),v(t)$ to :
\[\ddot u - \Lambda (t)u = 0\]
\begin{equation}\ddot v - \Lambda (t)v = 0\end{equation}
The time dependent Wronskian associated with these two solutions is
\begin{equation}W = {u_i}{{\dot v}_i} - {v_i}{{\dot u}_i} = {u_f}{{\dot v}_f} - {v_f}{{\dot u}_f}\end{equation}
where $i,f$ corresponds to the evaluation initial and final time slices. The solution can be normalized so that 
\begin{equation}{u_i}{v_f} - {u_f}{v_i} = 1\end{equation}
The final expression for the transition probability amplitude is then:
\begin{equation}K = \sqrt {\frac{W}{{2\pi i}}} \exp \left[ {\frac{i}{2}\left( {\left( {{u_i}{{\dot v}_f} - {v_i}{{\dot u}_f}} \right)q_f^2 + \left( {{u_f}{{\dot v}_i} - {v_f}{{\dot u}_i}} \right)q_i^2 - 2W{q_i}{q_f}} \right)} \right]\end{equation}
Now specializing to the case where where 
\begin{equation}\Lambda (t) =  - \frac{{{k^2}}}{{{t^2}}}\end{equation}
where we have defined:
\begin{equation}{k^2} =  - \frac{{\left( {{\beta ^2} - 2\alpha \beta } \right)}}{{{{\left( {2\alpha } \right)}^2}}}\end{equation}
The 3d Matrix theory time dependent deformation is related to one of the simplified cases studied in \cite{Craps:2010cn}. The two independent solutions are:
\[u = c{t^a}\]
\begin{equation}v = c{t^{1 - a}}\end{equation}
with
\begin{equation}a = \frac{1}{2}\left( {1 + \sqrt {4{k^2} + 1} } \right)\end{equation}
and 
\begin{equation}c = {\left( {t_i^at_f^{1 - a} - t_f^at_i^{1 - a}} \right)^{ - 1/2}}\end{equation}
The Wronskian is then
\begin{equation}W = {c^2}(1 - 2a)\end{equation}
This can be compared to the ordinary harmonic oscillator for which 
\[u(t) = \frac{{{\mathop{\rm cos}\nolimits} \omega t}}{{\sqrt {{\mathop{\rm sin}\nolimits} \omega T} }}\]
\begin{equation}v(t) = \frac{{{\mathop{\rm sin}\nolimits} \omega t}}{{\sqrt {{\mathop{\rm sin}\nolimits} \omega T} }}\end{equation}
and 
\begin{equation}W = \frac{\omega }{{\sin \omega T}}\end{equation}
so that:
\begin{equation}K = \sqrt {\frac{\omega }{{2\pi i\sin \omega T}}} \exp \left[ {\frac{{i\omega }}{{2\sin \omega T}}\left( {\cos \omega Tq_f^2 + \cos \omega Tq_i^2 - 2{q_i}{q_f}} \right)} \right]\end{equation}
with $t_f=T$ and $t_i=0$.

\subsection*{Gauss law}

The bosonic Matrix $X$ and it's canonical momentum can be expanded as:
\[X = {x_a}{\sigma ^a}\]
\begin{equation}P = {p_a}{\sigma ^a}\end{equation}
In the gauge $A_0=0$ the Gauss law constraint is given by:
\begin{equation}G=\left[ {X,P} \right] = 0\end{equation}
which in terms of the expansion coordinates becomes:
\begin{equation}{G_a=\varepsilon ^{abc}}{x_b}{p_c} = 0\end{equation}
And physical states obey:
\begin{equation}{G_a}\left| \psi  \right\rangle  = 0\end{equation}
It is important that the Gauss law constraint is imposed otherwise for example in time evolution spurious states will interfere with the evolution of the physical states. On the other hand if one has a physical state it will evolve into another physical state. This is because the Hamiltonian commutes with the Gauss law constraints. It obtain a physical states on can form the integral 
\begin{equation}\left| {{\psi _{phys}}} \right\rangle = \int {d\alpha {e^{i\alpha {G_a}}}\left| {{\psi _{extended}}} \right\rangle } \end{equation}
where ${\left| {{\psi _{extended}}} \right\rangle }$ is a state in the full extended Hilbert space. One can also add $\lambda G^2_a$ to the Hamiltonian as a penalty term. With $\lambda$ large the non gauge invariant states will obtain a large energy and will not interfere with the low lying physical states.

The time evolution of a physical state can be expressed as a double integral of the time evolution of of a state in extended Hilbert space as follows
\[\left\langle {{\psi _{phys}},f} \right|{T e^{ - i\int H(t)dt}}\left| {{\psi _{phys}},i} \right\rangle  = \left\langle {{\psi _{extended}},f} \right|\int {d\alpha '{e^{ - i\alpha' {G_a}}}} T{e^{ - i\int dt H(t)dt}}\int {d\alpha {e^{i\alpha {G_a}}}} \left| {{\psi _{extended}},i} \right\rangle \] \begin{equation}= \int {d\alpha 'd\alpha \left\langle {{\psi _{extended}},f} \right|{Te^{ - i\int (\alpha'{G_a'} - iH(t) + i\alpha {G_a})dt}}\left| {{\psi _{extended}},i} \right\rangle } \end{equation}
Thus one can add Gauss constraints to the Hamiltonian for time evolution of extended Hilbert space states to gain information about physical state evolution. Besides implementing a projection operator one can also solve the gauge constraint so that only physical sates contribute. Another method is to go to a diagonal basis in which the diagonal entries or eigenvalues obey Fermi statistics. This leads to a fermionic formulation of the noncritical M-theory which would also make a good starting point for quantum computation. This formulation has a great similarity to quantum chemistry as well as to approaches to 2d Yang Mills. Finally for large $N$ it may not be necessary to invoke the Gauss law constraint as the global $SU(N)$ symmetry may be sufficient \cite{Maldacena:2018vsr}
\cite{Pateloudis:2022oos}. 

In the supersymmetric 3d M-theory Matrix model the Hamiltonian and Gauss law constraint is given for $N=2$ as:
\[H(t) = \frac{1}{2}{p_a}{p_a} + \frac{1}{2}{\omega ^2}(t){x_a}{x_a} +g{\varepsilon _{abc}}\psi _a {x_b}{\psi _c}+\psi_a^\dag \psi_a\]
\begin{equation}{G_a} = {\varepsilon _{abc}}\left( {{x_b}{p_c} - i\psi _b^\dag {\psi _c}} \right)\end{equation}
In this case time evolution is similar to that of a time dependent version of the supersymmetric Harmonic oscillator. Instead of projecting the extended Hilbert space one can also work with a physical basis of states annihilated by the Gauss law constraints. This has the advantage of reducing the size of the Hilbert space and the number of qubits required for the EOH algorithm. For example for the bosonic Matrix model using three $4\times 4$ bosonic Matrices so $Nq=6$ there are $64$ states labeled by  $\left| {{n_x},{n_y},{n_z}} \right\rangle $ where $n_i$ are integers $0,1,2,3$. The physical basis states are given by:
\[\left| {{\psi _{phys0}}} \right\rangle  = \left| {0,0,0} \right\rangle \]
\[\left| {{\psi _{phys1}}} \right\rangle  = \left( {\sum\nolimits_{i = 0}^3 {A_i^\dag A_i^\dag } } \right)\left| {0,0,0} \right\rangle \]
\[\left| {{\psi _{phys2}}} \right\rangle  = {\left( {\sum\nolimits_{i = 0}^3 {A_i^\dag A_i^\dag } } \right)^2}\left| {0,0,0} \right\rangle \]
\begin{equation}\left| {{\psi _{phys3}}} \right\rangle  = {\left( {\sum\nolimits_{i = 0}^3 {A_i^\dag A_i^\dag } } \right)^3}\left| {0,0,0} \right\rangle \end{equation}
These states also have an interesting interpretation as closed string in toy models of $AdS/CFT$ \cite{Berenstein:2004kk}. Because there are only four physical states now only $N_{phys}=2$ qubits are required which speeds up the computations. Similar reductions are possible for the simulation of more complex supersymmetric theories \cite{Baskan:2019qsb}
\cite{Asano:2015eha}
\cite{Culver:2021rxo}
\cite{Schaich:2022duk}
\cite{Schaich:2022xgy}
\cite{Buividovich:2022udc}
\cite{Hubener:2014pfa}
\cite{Berenstein:2016zgj}
\cite{Kares:2004uk}
\cite{Wosiek:2002nm}
\cite{Korcyl:2009yh}
\cite{Korcyl:2010uq}
\cite{Korcyl:2011vr}
\cite{Campostrini:2002mr}
\cite{Hanada:2010rg}
\cite{Wosiek:2004yg}. 
\subsection*{Quantum computation}

The Evolution of Hamiltonian (EOH) quantum algorithm represented the time evolution operator $U(t) = e^{-i H t}$ on a quantum computer by representing $U(t)$ as a series of unitary gates though the Trotter approximation. In the algorithm one has control over to the number of time slices which can be used to increase the accuracy of the simulation but a the cost of greater circuit depth. 
Quantum computation of the evolution of time dependent Hamiltonians is an area of active research. See for example \cite{Baishya}
\cite{Lau}
\cite{Zhao:2022dma}
To compute the transition amplitude $K$ on a quantum compute one needs to construct the time ordered Unitary operator 
\begin{equation}U = T\left( {{e^{ - i\int_{{t_i}}^{{t_f}} {dtH(t)} }}} \right)\end{equation}
in terms of quantum gates and evaluate the represented quantum circuit. One approach is to break up the circuit into time intervals and evaluate.
\begin{equation}U = \prod\limits_{t = {t_i}}^{{t_f}} {{e^{ - i\Delta tH(t)}}} \end{equation}
For our computations we took $\Delta t = .04$ and a total of 100 times slices using the Trotter approximation. We use the fact that for the bosonic Matrix model $K$ factorizes into three identical factors so we only need to simulate one factor. Figure shows a comparison for the time evolution for 2 qubits for the ordinary simple harmonic oscillator versus the time dependent harmonic oscillator used in the Matrix big bang cosmology. One can see some notable differences at late times where the frequency decreases for the Matrix big bang solution. The results of the quantum EOH computation are in table 2. The results are highly accurate using the state vector simulator in IBM QISKIT. The code can be easily run on quantum hardware but current quantum computers are susceptible to noise and short coherence times. Also for the quantum hardware for the Hamiltonian evolution one needs to use quantum state tomography to recover the full quantum state rather than just the magnitude squared of the amplitude.
\begin{figure}[!htb]
\centering
\minipage{0.5\textwidth}
  \includegraphics[width=\linewidth]{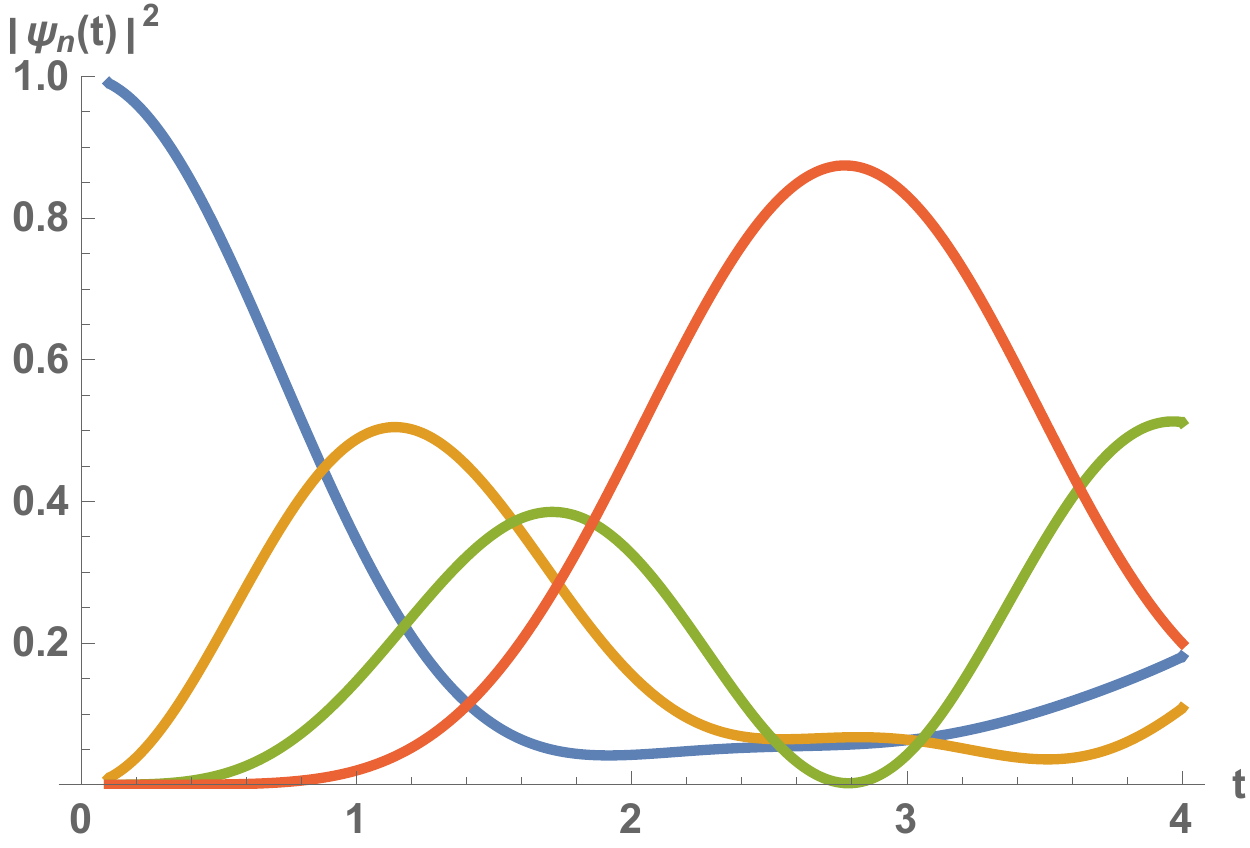}
\endminipage\hfill
\minipage{0.5\textwidth}
  \includegraphics[width=\linewidth]{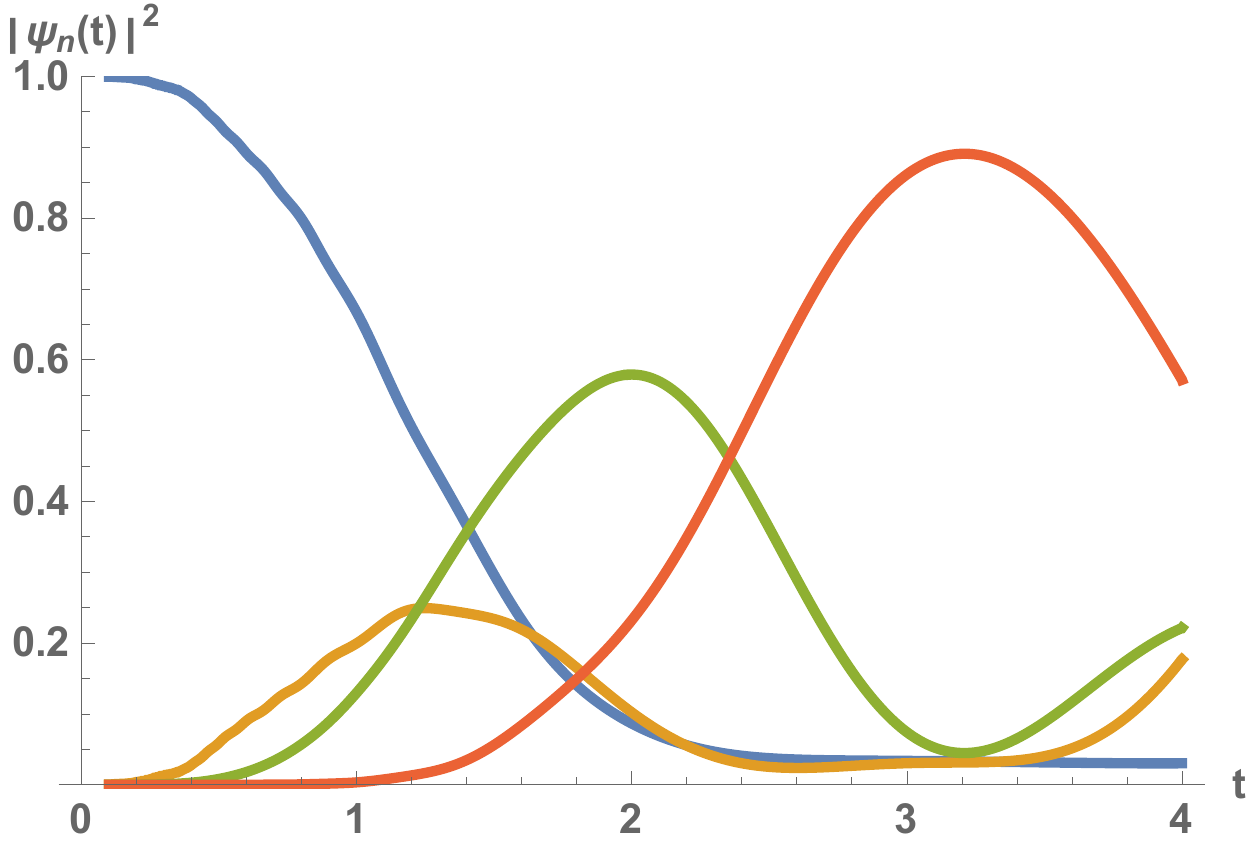}
\endminipage\hfill
\caption{Time dependence of norm squared for state vectors for 2 qubit 4 state Evolution of Hamiltonian computation for ordinary Simple Harmonic Oscillator (left) Time dependent Simple Harmonics Oscillator used to describe Matrix cosmology (right). 
}
\end{figure}
\begin{figure}
\centering
  \includegraphics[width = 1 \linewidth]{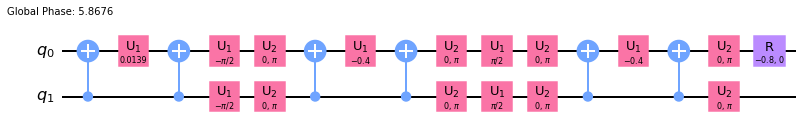}
  \caption{Quantum circuit for the EOH quantum algorithm for one time slice of the Matrix big bang time evolution using 2 qubits and 5 Pauli terms and the Trotter approximation.}
  \label{fig:Radion Potential}
\end{figure}
\begin{table}[ht]
\centering
\begin{tabular}{|l|l|l|l|l|l|l|}
\hline
3d M-theory  & basis    & Qubits  &  Paulis  & Exact Discrete  & EOH Result \\ \hline
$\psi_1(t=4)$  & fd & 2 & 5  & $0.1234653
  + 0.12267727 i$ &  $0.12346547+0.12267706 i$   \\ \hline
$\psi_2(t=4$  & fd & 2 & 5  & $0.07651227 - 0.03372487 i$ &  $0.07651222-0.03372502 i$   \\ \hline
$\psi_3(t=4)$  & fd & 2 & 5  & $-0.10583990 - 0.6162668 i$ &  $-0.10584101-0.61626668 i$   \\ \hline
$\psi_4(t=4)$  & fd & 2 & 5  & $-0.58216811 + 0.48250234 i$ &  $-0.58216728+0.48250332j$   \\ \hline

\end{tabular}
\caption{\label{tab:BasisCompare}  EOH results for   the Hamiltonian for 3D M-theory Matrix cosmology using the time dependent quantum oscillator and finite difference  basis for an initial time of .1 and final time of 4. 
}
\end{table}

\section{BRST quantization}

Beside the traditional Hamiltonian quantization considered above one can also consisder more modern quantization methods such as BRST (Becchi, Rouet, Stora and Tyutin) quantization \cite{Barnich:2000zw} from the point of view of quantum computing. As with fermions, ghosts can be introduced using Grassmann variables which can be computationally intensive to simulate on a classical computer. Thus BRST quantization with quantum computers may be an area of quantum advantage over classical computers. BRST quantization typically involves the introduction of ghost fields and the definition of the BRST charge using the constraints of the system which annihilates physical states. We describe the BRST quantization of the SU(2) 3d M-theory Matrix Model following the treatment of \cite{Fuster:2005js} for the Unitary and Lorenz gauge. 

\subsection*{Unitary gauge}

The Unitary gauge for the SU(2) 3d M-theory Matrix Model is given by the condition:
\begin{equation}{A_0} = 0\end{equation}
The Lagrangian of the bosonic Matrix model then becomes:
\begin{equation}{L_{eff}} = \frac{1}{2}\dot X_a^2 + {b_a}{{\dot c}_a}\end{equation}
where we have introduced  $b,c$ ghost fields. The Hamiltonian is simply:
\begin{equation}{H_{eff}} = \frac{1}{2}P_a^2\end{equation}
and the BRST charge is given by:
\begin{equation}\Omega  = {c_a}{G_a} - i\frac{g}{2}{\varepsilon ^{abc}}{c_a}{c_b}{b_c}\end{equation}
with the gauge constraints given by:
\begin{equation}{G_a} = g{\varepsilon ^{abc}}{X_b}{P_c}\end{equation}
The BRST charge obeys the nilpotency condition:
\begin{equation}{\Omega ^2} = 0\end{equation}
and annihillates physical states from the equation:
\begin{equation}\Omega \Psi  = 0\end{equation}
The BRST charge can also be used to define an equivalency class of physical states given by: 
\begin{equation}\Psi  \sim \Psi ' + \Omega \Lambda \end{equation}

\subsection*{Lorenz gauge}
The BRST treatment of the Lorenz gauge is more complicated. The Lorenz gauge condition is:
\begin{equation}{{\dot A}_0} = 0\end{equation}
and the Lagrangian is:
\begin{equation}{L_{Lorenz}} = \frac{1}{2}{\left( {{D_0}{X_a}} \right)^2} + \frac{1}{2}{\left( {{{\dot A}_0}} \right)^2} - i{{\dot b}_a}{\left( {{D_0}c} \right)_a}\end{equation}
The Hamiltonian is given by:
\begin{equation}{H_{Lorenz}} = \frac{1}{2}{({P_a} + g{\varepsilon _{abc}}A_0^b{X^c})^2} + \frac{1}{2}{\left( {P_0^a} \right)^2} - \frac{{{g^2}}}{2}{\left( {{\varepsilon _{abc}}A_0^b{X^c}} \right)^2} - i{\left( {{D_0}c} \right)^a}{{\dot b}_a} + g{\varepsilon _{abc}}A_0^b{c^c}{{\dot b}_a}\end{equation}
The BRST charge for the Lorenz gauge is:
\begin{equation}\Omega  = {G_a}{c_a} - i\frac{g}{2}{\varepsilon _{abc}}{c_a}{c_b}{{\dot b}_c} - P_0^a{\left( {{D_0}c} \right)^a}\end{equation}

\subsection*{Quantum computing}
In the Unitary gauge the quantum state depends on matrix valued $X$ and $c$ ghost fields given by:
\begin{equation}\Psi ({X_a},{c_a})\end{equation}
In the Lorenz gauge the quantum state depends also on the $b$ ghost field and zero component of the vector potential and is given by:
\begin{equation}\Psi ({X_a},{A_0},{c_a},{b_a})\end{equation}
Defining the BRST coordinates $Z$ as:
\begin{equation}Z = \left( {{X_a},{c_a}} \right)\end{equation}
for the Unitary gauge and
\begin{equation}Z = \left( {{X_a},{A_0},{c_a},{b_a}} \right)\end{equation}
for the Lorenz gauge. The BRST propagator \cite{vanHolten:1995ds} is defined by
\begin{equation}K(Z,Z'|T) = \left\langle Z \right|T{e^{ - i\int_0^T {{H_{eff}}(t)dt} }}\left| {Z'} \right\rangle \end{equation}
Representing three $X_a$ fields as $4\times 4$ matrices and the $b$ and $c$ ghost fields by $2\times 2$ matrices with require $9$ qubits for the Unitary gauge and $12$ qubits for the Lorenz gauge. If instead representing three $X_a$ fields as $3\times 3$ matrices and the $b$ and $c$ ghost fields by $2\times 2$ matrices with require $8$ qubits for the Unitary gauge and $11$ qubits for the Lorenz gauge.

The BRST charge is not Hermitian and thus not suitable for VQE quantum computation. However one can form the BRST Laplacian defined by:
\begin{equation}
Q_L = {\Omega ^\dag }\Omega    
\end{equation}
which is Hermitian. The VQE results for the BRST Laplacian in the Unitary gauge using $X_a$ fields as $3\times 3$ matrices and 8 qubits are given in figure 6 and table 3 with highly accurate results using the quantum simulator with no noise.
\begin{figure}
\centering
  \includegraphics[width = 1 \linewidth]{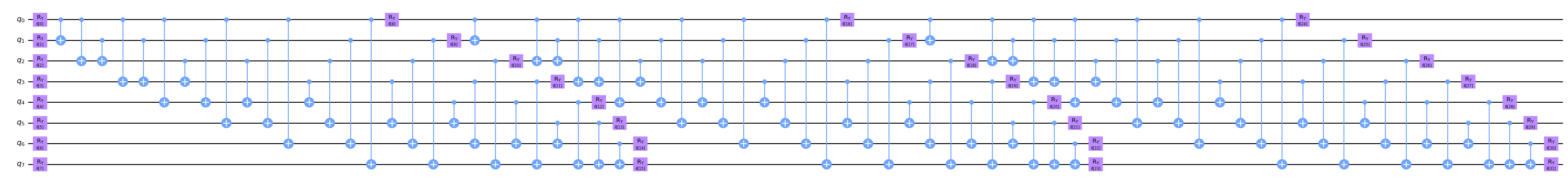}
  \caption{Variational ansatz represented as a 8 qubit quantum circuit for the quantum VQE computation of the BRST Laplacian $Q_L$ for the 3d M-theory Matrix Model using the oscillator basis. The ansatz uses  parametrized $R_y$  gates with a depth of three and thirty two parameters }
  \label{fig:Radion Potential}
\end{figure}
\begin{figure}
\centering
  \includegraphics[width = .4 \linewidth]{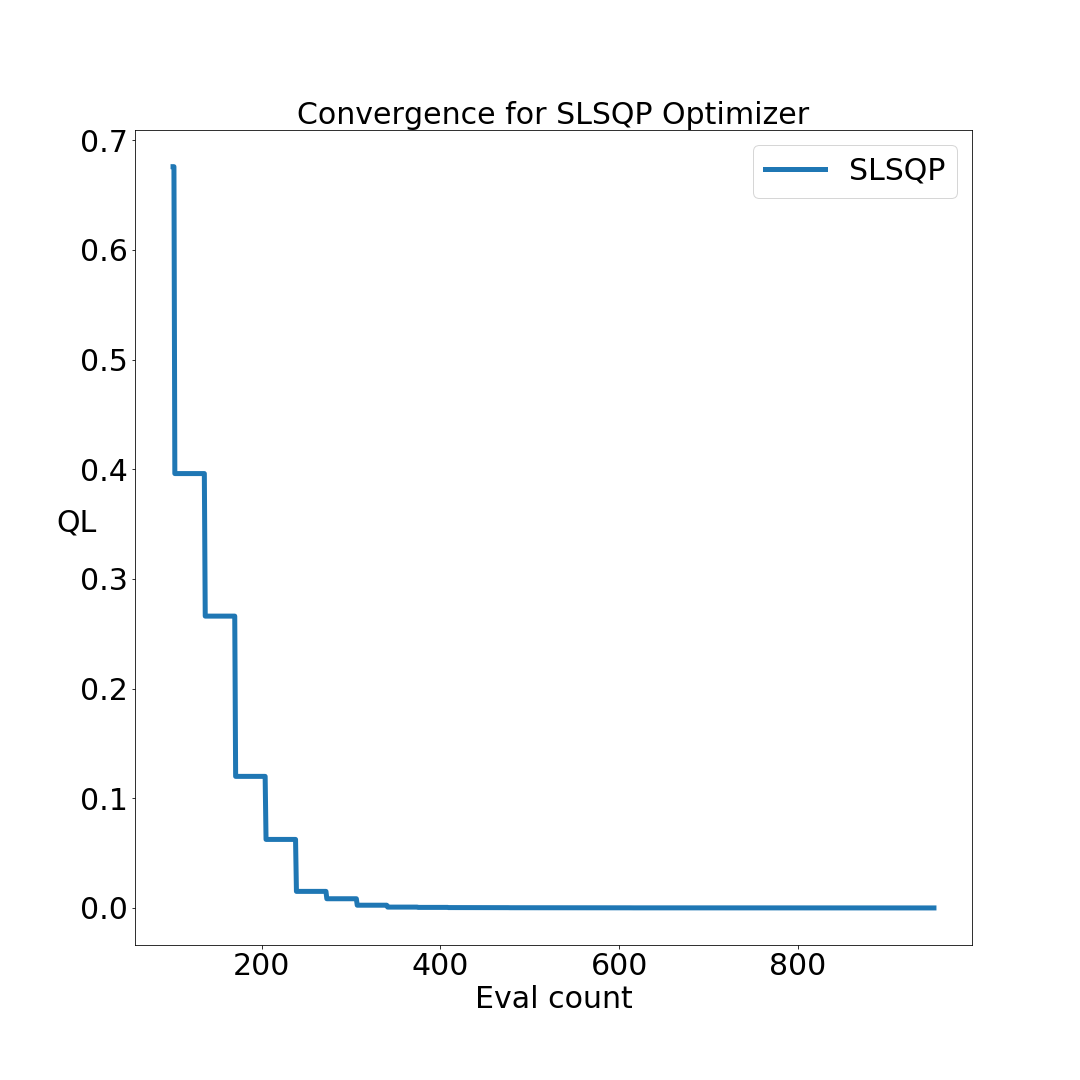}
  \caption{Convergence graph for the quantum VQE computation of the BRST Laplacian $Q_L$ for the 3d M-theory Matrix Model using the oscillator basis.   The VQE result was obtained using the State-vector simulator with no noise and the Sequential Least SQuares Programming (SLSQP)  optimizer.}
  \label{fig:Radion Potential}
\end{figure}

\begin{table}[ht]
\centering
\begin{tabular}{|l|l|l|l|l|l|l|}
\hline
3d M-theory  & basis    & Qubits  &  Paulis  & Exact  & Exact Discrete & VQE Result \\ \hline
$Q_L$  & osc & 8 & 1494  & $0.0$ &  $-4.5554\times 10^{-15}$ & $1.3511\times 10^{-5}$ \\ \hline

\end{tabular}
\caption{\label{tab:BasisCompare}  VQE results for the lowest eigenvalue of the Hamiltonian for 3D M-theory using the oscillator and finite difference  basis. 
}
\end{table}

\section{Conclusion}

In this paper we studied the application of quantum computing to the Matrix model associated with 3d M-theory which is simpler than the BFSS or BMN Matrix model of 11d M-theory. We applied the VQE quantum algorithm in the finite difference and oscillator basis and compared the results. We also applied the EOH quantum algorithm to the time dependent Matrix model associated with a Matrix Big Bang space-time in 3d M-theory taken as a subspace of the 11d solution and discussed ways to implement the Gauss law constraint within the EOH algorithm and using BRST methods. As quantum computers become more powerful and more accurate it will be interesting to extend these techniques to 11d M-theory and other space-times. Also there is a noncritical version of 4d M-theory that could be implemented on a quantum computer as an intermediate step \cite{Polacek:2014cva}. Finally although it is interesting that one can explore singular space-times on quantum computers using Matrix models these still represent specific space-times. It would be better to have a formulation of Matrix theory which didn't rely on a specific background such as a background independent approach. One of the earliest such approaches involve the Wheeler-DeWitt equation \cite{DeWitt:1967yk}\cite{Rovelli:2015gwa} and work has been done connecting the Matrix model on arbitrary backgrounds to the Wheeler-DeWitt equation \cite{Matsuo:2008yd}\cite{Lifschytz:2000bj}. It would be interesting to implement these  Matrix model approaches to quantum gravity using quantum computing as a way of realizing  quantum gravity on a quantum computer  \cite{Beane:2012rz}\cite{Lloyd}\cite{Lloyd:2005js}.

\section*{Acknowledgements}
We would like to to thank Tristen White for discussions about simulating time dependent Hamiltonians on quantum computers and Mohammad Hassan for help with quantum computing software.

\end{document}